# Performance Optimization on Practical Quantum Random Number Generators: Modification on Min-entropy Evaluation and Acceleration on Post Processing

Zehao (Andy) Zhao

Princeton International School of Mathematics and Science

Princeton, NJ, USA

under the direction of

Professor Xiongfeng Ma

and

Mr. Hongyi Zhou

Institute for Interdisciplinary Information Sciences

Tsinghua University

## Abstract

Quantum random number generation is a technique to generate random numbers by extracting randomness from specific quantum processes. As for practical random number generators, they are required not only to have no information leakage but also have a high speed at generating random sequences. In this paper, we consider the generators based on laser phase noise and propose a method to modify the estimation of min-entropy, which can guarantee no information leakage to the eavesdropper. We also accelerate post processing based on Toeplitz matrix with Fast Fourier Transformation, reducing its time complexity to O(nlogn). Furthermore, we discuss the influence on post processing speed by block length and find a proper block length to process a fixed-length raw sequence.

## Summary

Quantum random number generation is a technique to generate true random numbers. As for practical random number generators, they are required not only to have no information leakage but also have a high speed at generating random sequences. In this paper, we propose a method guarantee no information leakage to the eavesdropper. We also accelerate post processing and find a minimum time spent. The results after post processing successfully pass the statistical tests of randomness.

# I. Introduction

Random numbers are widely used in a variety of practical applications, such as cryptography[1-4], scientific simulations[5-8], economy[9], etc. Commonly, pseudo random numbers are used in such fields, being generated from math algorithms[10-12] or classical physics procedures[13-18]. However, only random numbers generated from specific quantum processes are considered to be true random numbers, whose characteristics include unpredictability, irreproducibility and non-autocorrelation. Thus, quantum random number generator (QRNG) is an advanced generating facility.

There are different schemes to achieve QRNG with current technology, especially with optical techniques. For instance, detecting photons' arriving time[19-21], photons' distribution of quantum state[22-23], laser's phase noise[24-26] or the fluctuation of vacuum state[27-29] are all examples of technologies that depend on some form of QRNG. The signals we detect usually can be described with particular equations and follow specific mathematical distributions. Hence, the amount of randomness coming from quantum processes can be calculated by min-entropy theory with the help of these distributions[30-31]. Nevertheless, in practice, the signals, as well as classical noises, hardly follow the distributions perfectly. This may be aroused by noises, statistic fluctuations or even attacks from an eavesdropper, and leads to an overestimation of min-entropy and information leakage to him[32-33].

In this paper, we propose a method to modify the calculation of min-entropy when the data we get from measurement does not follow some distribution perfectly. We prove why this may lead to information leakage, or insecurity in the same sense, and how to modify under this situation by using the concept of statistic distance.

The raw data directed from measurement may follow a specific distribution; meanwhile, uniformly distributed random numbers are most widely used, so we also realize the conversion from an arbitrary-distributed raw number sequence to a uniform-distributed random number sequence with Toeplitz matrix in post processing. The algorithm is based on matrix multiplication, while we accelerate it by adopting fast Fourier transform (FFT); another method employing modified Toeplitz matrix requires even fewer seed length. Furthermore, a comparison will be made between the calculation durations when choosing different block sizes, to provide a reference to determine proper block size in post processing.

## II. *Evaluation of min-entropy and modification under possible attack*

### *A. Background of min-entropy*

Quantum random number generators (QRNG) generates random number sequences by extracting the randomness from specific quantum processes or quantum noises, or quantum randomness sources, such as the arriving time of photons[19-21], the photon distribution of optical states[22-23], laser phase noise[24-26], laser polarization noise[36], vacuum noise[27-29], Raman scattering[34-35], etc. Commonly, these processes or noises can be described with certain equations, so the signals detected can also be described with these equations and follow specific statistical distributions. For example, for a coherent state of amplitude $\alpha$, the probability of detecting zero photon is

$$P(n=0) = e^{-|\alpha|^2},$$

which follows an exponential distribution, while the probability of detecting one or more photons is

$$P(n \geq 1) = 1 - e^{-|\alpha|^2}.$$

The 'amount' of secure randomness extracted from these quantum sources can be calculated with min-entropy[30-31], considering there might be an eavesdropper during the generation of random numbers. As for a variable $x$ whose probability distribution is $P(x)$, its min-entropy $H_{min}(x)$ is

$$H_{min}(x) = -\log_2 \left[\max_{x \in A} P(x)\right],$$

where $A$ is the domain of $x$. $H_{min}(x)$ represents a lower bound of all entropy corresponding to the probability of guessing at the first attempt by the eavesdropper, because his optimized strategy is to guess the number with the highest probability of $P(x)$ (Fig. 1).

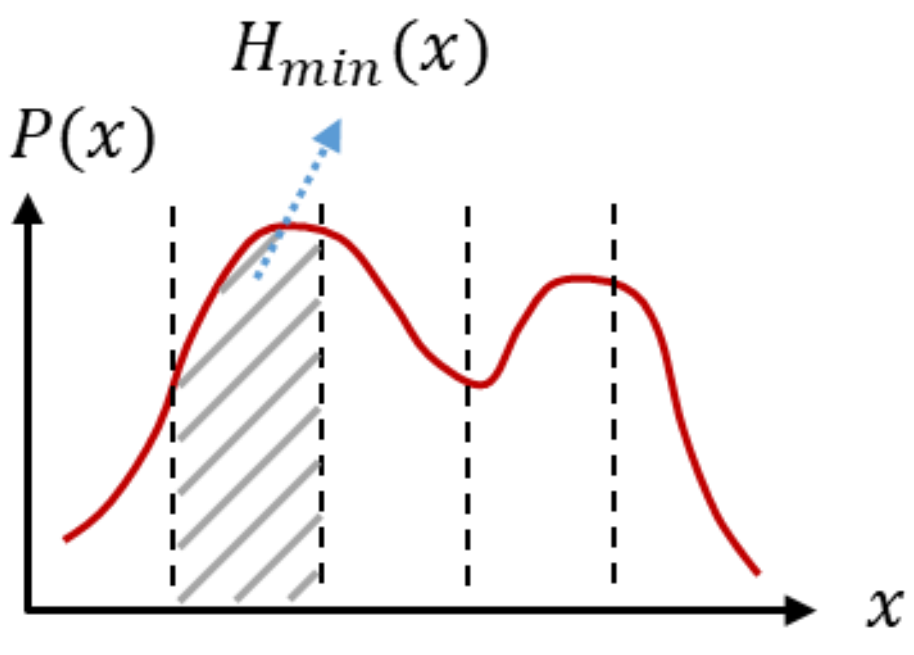

Fig.1 Calculation of min-entropy.

For example, if the probability of $P(x=1)$ is the biggest for variable $x$, the eavesdropper will have the biggest probability to guess $x$ correctly when he guesses $x = 1$ every time. As a result, the min-entropy is determined by $P(x = 1)$ to ensure the eavesdropper gets no information of the random numbers.

Any arbitrary distribution with min-entropy $H_{min}(x) = k$ can be written as a convex combination of distributions that are uniform for $k$ bits. This gives a critical interpretation of min-entropy as the number of secure uniform bits that can be extracted from this given distribution. Secure here means the there is no information leakage, which contains some properties of the distribution, to the eavesdropper. Thus, the eavesdropper cannot guess the final bits with a correct probability higher than 50%.

### B. *Attack on practical QRNG based on laser phase noise*

In a practical QRNG, randomness is extracted from the quantum source accompanied with classical noises and fluctuations, such as electrical noise and mechanical vibration. To ensure that the final random sequences to be secure, we usually assume these noises are all caused by the eavesdropper, and he has the ability to control these noises totally. Whatever quantum noises we detect, it has to be converted to classical signals by practical measurement devices. For instance, when measuring the laser's phase noise or polarization noise, the generator actually measures the intensity of laser and convert it into electric signals with the help of electric devices. As a result, it is inevitable to introduce classical noises in the measurement results. Therefore, the influences of these classical noises should be subtracted from calculation of min-entropy, as

$$H_{min}^{q}(x) = H_{min}^{q+c}(x) - H'^{c}(x'),$$

where q refers to 'quantum' and c refers to 'classical.'

Normally, not only quantum noises but also classical noises are assumed to follow specific distributions perfectly. For instance, when measuring the phase noise of a laser source, the distribution of phase noise, as well as classical electric noises, follow the Gaussian distribution [24,37]. The variances of quantum+classical noise $\sigma_{q+c}^2$ and classical noise $\sigma_c^2$ can be measured with the laser on and off respectively. So the variance of quantum noise itself, $\sigma_q^2$, is

$$\sigma_q^2 = \sigma_{q+c}^2 - \sigma_c^2.$$

With $\sigma_q^2$, the noise's Gaussian distribution is determined, and its min-entropy can be calculated.

However, a statistical distribution can be much different from a Gaussian distribution even though they have the same variance. This may be caused by the following factors:

unexpected noises in the electric devices and optical components; approximations introduced by the limited sampling precision; fluctuations caused by finite-key-size effect.

More seriously, if the eavesdropper can control the classical noises or even affect the measuring of quantum noises, which is achievable with current technologies, he can make the actual signals follow a non-Gaussian distribution but have the same variance. Thus, the generator will get a different min-entropy $H_{min}(x')$ instead of actual min-entropy $H_{min}(x)$ when he does not notice the existence of eavesdropping. In this situation, the mutual information between the generator G and the eavesdropper E is,

$$I(G:E) = H_{min}(x) - H_{min}(x').$$

If $I(G:E) > 0$, the eavesdropper can get non-zero information of the random number sequences meaning the eavesdropper knows some properties of the random number sequences (such as some properties of their statistic distribution $P(x)$), so he can guess the number correctly with a higher probability with the help of this information, compared with a totally random guess. In this situation, the generator overestimates its min-entropy, leading to information leakage to the eavesdropper and the insecurity of QRNG.

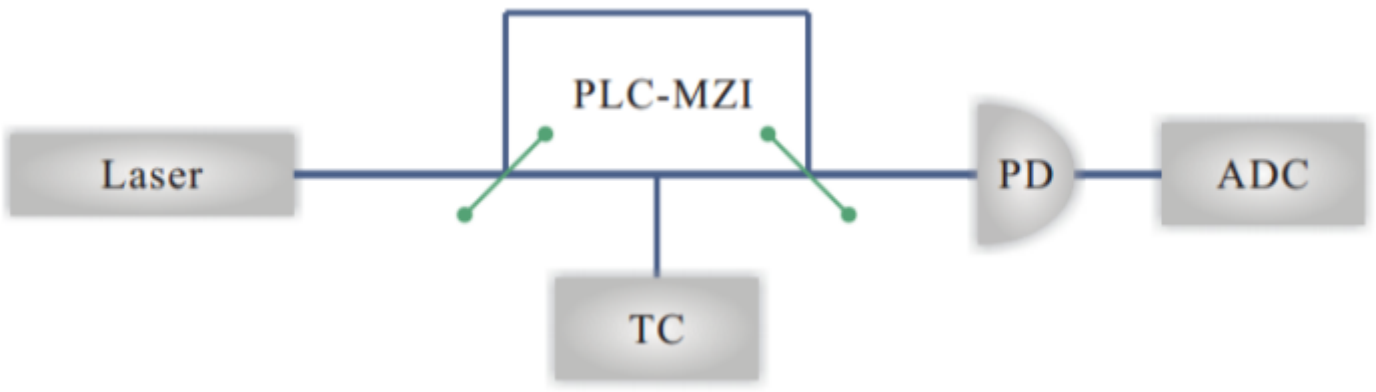

Fig. 2 Experiment setup of quantum random number generation from laser's phase noise. PLC-MZI: planar light wave circuit Mach-Zehnder interferometer; TC: temperature controller, used to stabilize the delay difference of PLC-MZI; PD: photodiode detector; ADC: analog-to-digital converter. This figure is excerpted from [24].

Here we propose an attack method for the eavesdropper to ensure $I(G:E) > 0$. When measuring the phase noise, the experiment setup is shown in Fig. 2, in which the photodiode detector (PD) converts laser intensity $I$ into electric voltage $V$. The PD's coefficient of converting light intensity into voltage $\alpha$ is constant, that is

$$V = \alpha_0 I.$$

But when the eavesdropper replaces the coefficient with a variable $\alpha(I)$ which is related to light intensity $I$ instead of a constant $\alpha_0$, and

$$\alpha(I) = \alpha_e\left(1 - \frac{I}{\beta_e I_{max}}\right),$$

he can change $V$ into

$$V' = \alpha(I)I.$$

By choosing proper parameters, $V'$ may have a non-Gaussian distribution but same variance and mean value as $V$.

Here we set $I \sim N(0,1), \alpha_0 = 1, \alpha_e = 0.99, \beta_e = 2$, and $I_{max} = 5$. Then the mean value of $V$ and $V'$ are both 0 while the variances of them are both 1, and the distribution is shown in Fig. 3, from which we can see the two figures have a significant 'distance' between them, leading to a difference between their min-entropy.

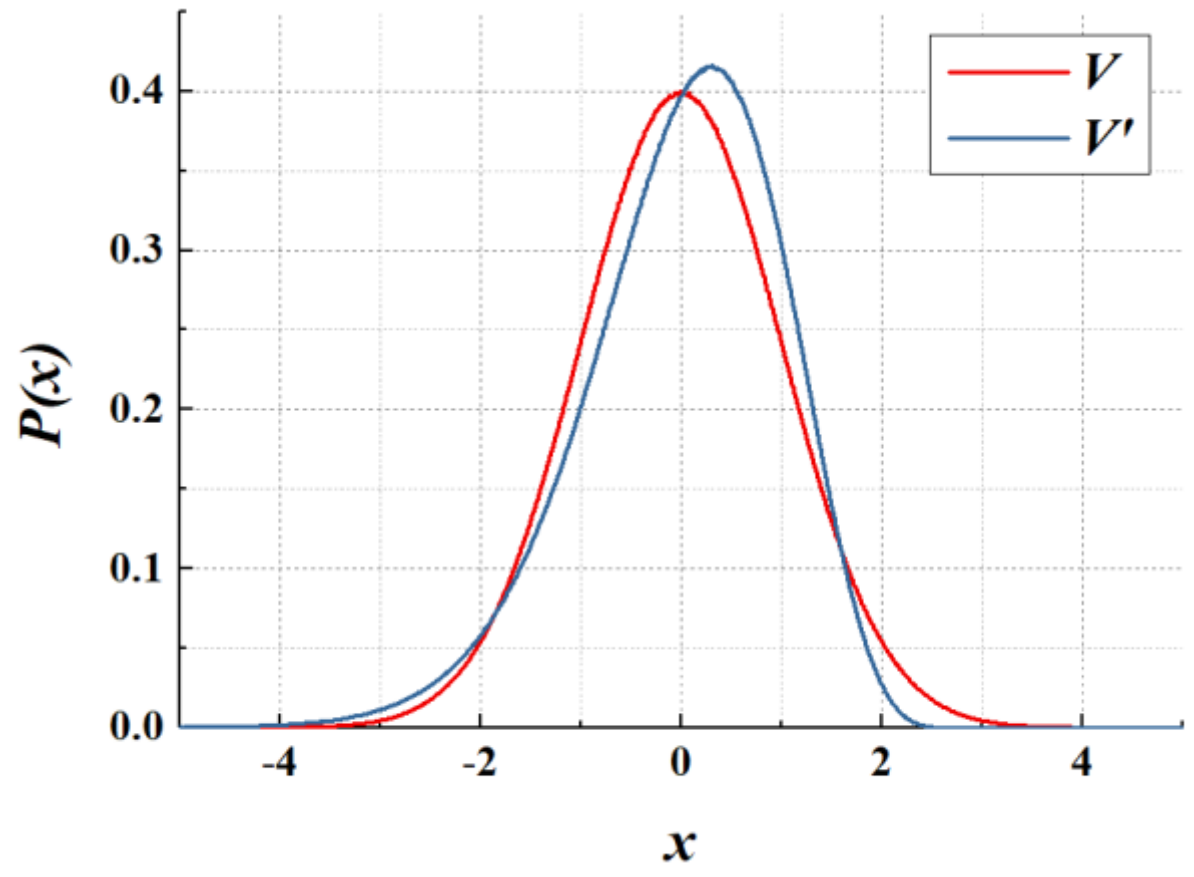

Fig. 3 Statistic distribution of $V$ and $V'$.

We assume the generator implements a 5-bit sampling, and the sampling range is -5 to 5. It can be seen apparently from Fig. 3 that when $x \in (0,1)$, $V'$ has a higher probability thus a lower min-entropy than $V$. So we have

$$H_{min}(V) \approx 2.99,$$
$$H'_{min}(V') \approx 2.94.$$

The mutual information

$$I(G{:}E) = H_{min}(V) - H'_{min}(V') = 0.05,$$

and the eavesdropper can get almost 1.7% information of the random sequences. This may seem to be little information, but can bring in catastrophes in a cryptosystem using random numbers. This information leakage can be larger when the eavesdropper adopts proper parameters to implement attacks, or when the original statistic distribution is not Gaussian, for example, when measuring the decay time of radiative atoms[38-40]. As a result, we must modify the calculation of min-entropy in QRNG to account for the possible attacks if the statistic distributions of raw data do not match specific distributions perfectly.

### *C. Modification under the possible attacks from the eavesdropper*

In practical QRNG devices, the statistical distribution of raw data may not follow a specific distribution perfectly, caused by parameter fluctuation, finite-size-effect, or even attack from the eavesdropper. In this situation, the min-entropy of the distribution can also change as shown in Sec. II. B. As a result, we need to modify the calculation of min-entropy in the case of information leakage to the eavesdropper, or $I(G\colon E) > 0$.

Here we explain why modification is necessary and how we do it. When calculating $H_{min}(V)$ with original distribution, we assume that there is no eavesdropper and the distribution of signals is what we exactly expect. However, in actual experiment, the eavesdropper may always exist and attack the devices without being noticed. So we should have a weaker assumption than before that attacking or eavesdropper always exists and the distribution is not what exactly we desire. Under this assumption, we should modify the original $H_{min}(V)$ with the data we actually get to ensure the generator's security. A weak assumption means stronger security in practice.

In this part, we give out a modification method in practical QRNG. According to the example in Sec. II, the mutual information between the generator and the eavesdropper, is

$$I(G\colon E) = H_{min}(x) - H'_{min}(x).$$

$H_{min}(x)$ is the min-entropy of the ideal distribution $P(x)$ while $H'_{min}(x)$ is the min-entropy of the actual distribution $P'(x)$, different from $P(x)$.As

$$H_{min}(x) = -\log_2\left[\max_{x\in A} P(x)\right],$$

we have ($U$ is the range of $x$ where we calculation min-entropy)

$$\begin{aligned}\Delta H = I(G\colon E) = H_{min}(x) - H'_{min}(x) &= \log_2\left[\max_{x\in A} P'(x)\right] - \log_2\left[\max_{x\in A} P(x)\right]\\ &= \log_2\left[\sum_{x\in U} P'(x)\right] - \log_2\left[\sum_{x\in U} P(x)\right] = \log_2\left[\frac{\sum_{x\in U} P'(x)}{\sum_{x\in U} P(x)}\right]\\ &= \log_2\left[\frac{\sum_{x\in U}(P'(x) - P(x))}{\sum_{x\in U} P(x)} + 1\right].\end{aligned}$$

$P'(x)$ here is used to describe an finite-size random sequence, but since we can only get a finite-size sequence in practice, we cannot always get accurate $P'(x)$. As a result, we should amplify $\Delta H$ to ensure the QRNG's security when considering statistic fluctuation caused by the finite-size effect.

As for $n$-bit sampling, we have

$$\Delta H = \log_2\left[\frac{\sum_{x\in U}(P'(x)-P(x))}{\sum_{x\in U}P(x)}+1\right] \le \log_2\left[\frac{\sum_{x\in U}|P'(x)-P(x)|}{\sum_{x\in U}P(x)}+1\right]$$
$$= \log_2\left[\frac{\sum_{x\in U}|P'(x)-P(x)|}{2^{-H_{min}}}+1\right] \le \log_2\left[\frac{2d_U(P',P)}{2^{-n}}+1\right] = \log_2\left[\frac{d_U(P',P)}{2^{-n-1}}+1\right]$$
$$= \Delta H_m,$$

where $d_U(P_1,P_2) = \frac{1}{2}\sum_{x\in U}|P_1(x=u)-P_2(x=u)|$ is the definition of statistic distance between two distribution $P_1(x)$ and $P_2(x)$[45].

$\Delta H_m$ is the modification term of $H_{min}(x)$, so the min-entropy after modification, or the amount of secure information we can extract from the random sequence, is

$$H_{min}^{modify}(x) = H_{min}(x) - \Delta H_m.$$

We re-calculate the min-entropy with this modification in the example of phase noise measuring in Sec. II A, and when the eavesdropper attacks the generator by the method above, we have

$$\Delta H_m = 0.10,$$
$$H_{min}^{modify}(V) = H_{min}(V) - \Delta H_m = 2.89 < 2.94 = H'_{min}(V').$$

The min-entropy we get is smaller than the actual mean entropy, so there is no information leakage to the eavesdropper, or $I(G{:}E) \le 0$. Although the generator extracts a little bit less random bits from each sampling after modification, the eavesdropper cannot get information from them. Thus, he does know any information of the properties of the distribution of actual signal, and cannot guess the random bits with a correct probability higher than 50%. This guarantees the security of pracital generators under possible attacks.

In practical a QRNG, we can select some bits randomly from the raw sequence and calculate the modified min-entropy $H_{min}^{modify}(x)$ of them, thus simplifying the process of modification.

## III. Acceleration on Post Processing in practical QRNG

### *A. Background of post processing*

Post processing is the procedure to process raw data, or the data got directly from digital sampling. On the one hand, the raw data contains 'unsecure' information. When implementing $N$-bit sampling, the lower bound of information can be extracted from each sampling is $k = H_{min}(x)$ if we require no information leakage to the eavesdropper. So post processing is adopted to compress the amount of data from $N$ to $k$ each sampling. On the other hand, the raw data usually does not follow the uniform distribution, which is required in most cases of random numbers, so post processing is adopted to change arbitrary-distributed random number sequences into "almost" uniform-distributed random numbers.

There are different ways of post processing, such as Von Neumann Rectification[41-43], Exclusive OR (XOR)[44], Least Significant Bit (LSB) method, etc. However, they are used to process raw bits under certain conditions. According to Sec. II, the raw data from sampling may follow arbitrary distribution, thus proper post processing algorithm should be employed in practical QRNG.

### *B. Acceleration on Post processing with Fast Fourier Transform*

Here we use Toeplitz matrix to achieve post processing. Such a Toeplitz matrix is a class of randomness extractor and can be used to get an "almost" random sequence[45]. It has been proved that by multiplying a kind of special Toeplitz matrix, an arbitrary-distributed random number sequence can be converted into a uniform-distributed random number sequence[46-48], which is a crucial property for practical QRNG, for uniform-distributed random numbers are most widely used.

A Toeplitz matrix transforms an $n$-bit-long sequence into an $m$-bit-long sequence, where

$$\boldsymbol{m} = \boldsymbol{n} \times \frac{\boldsymbol{H_{min}}}{\boldsymbol{N}} - \boldsymbol{2}\log_2\left(\frac{\boldsymbol{1}}{\boldsymbol{\epsilon}}\right).$$

We can see $m$ is determined by digits of sampling $N$, min-entropy $H_{min}$ and the security parameter $\epsilon$ according to leftover Hash lemma[47], indicating how close the distribution of final keys is to uniform distribution. And the extraction ratio of this matrix is $\gamma$, where $\gamma = m/n$.

In post processing, we construct an $m \times n$ Toeplitz matrix in the following form[49]: In every diagonal, the elements are the same. That is, $a_{i,j} = a_{i+1,j+1}$ $(1 \leq i \leq m-1, 1 \leq j \leq n-1)$. So Toeplitz matrix $T$ is,

$$m\text{ bits}\left\{\begin{pmatrix} k_1 \\ k_2 \\ \vdots \\ k_{m-1} \\ k_m \end{pmatrix}\right.$$

$$= \begin{pmatrix} a_n & a_{n-1} & a_{n-2} & & a_3 & a_2 & a_1 \\ a_{n+1} & a_n & a_{n-1} & \cdots & a_{n+1} & a_3 & a_2 \\ a_{n+2} & a_{n+1} & a_n & & a_5 & a_4 & a_3 \\ & \vdots & & \ddots & & \vdots & \\ a_{n+m-3} & a_{n+m-4} & a_{n+m-5} & & a_m & a_{m-1} & a_{m-2} \\ a_{n+m-2} & a_{n+m-3} & a_{n+m-4} & \cdots & a_{m+1} & a_m & a_{m-1} \\ a_{n+m-1} & a_{n+m-2} & a_{n+m-3} & & a_{m+2} & a_{m+1} & a_m \end{pmatrix} \left.\begin{pmatrix} r_1 \\ r_2 \\ \vdots \\ r_{n-1} \\ r_n \end{pmatrix}\right\} n\text{ bits},$$

where $r$ refers to raw data and $k$ refers to final keys.

This method should be simplified for calculation in a computer, since it is difficult to store a huge matrix when $m$ and $n$ are too large. The time complexity of direct computation is $O(mn)$. Though is a polynomial time complexity, it is still not efficient enough.

To avoid such an inefficient method, we extend Toeplitz matrix to a circulant matrix which is determined by its first column. We define the circulant matrix as follow.

**Definition 1:** A circulant matrix is a matrix which the last element of the $j$-th column is the first element of the $j+1$-th column when the first column is determined. That is,

$$A = \begin{pmatrix} a_1 & a_{n+m-1} & a_{n+m-2} & & & a_4 & a_3 & a_2 \\ a_2 & a_1 & a_{n+m-1} & \cdots & & a_5 & a_4 & a_3 \\ a_3 & a_2 & a_1 & & & a_6 & a_5 & a_4 \\ & \vdots & & \ddots & & & \vdots & \\ a_{n+m-3} & a_{n+m-4} & a_{n+m-5} & & a_1 & a_{n+m-1} & & a_{n+m-2} \\ a_{n+m-2} & a_{n+m-3} & a_{n+m-4} & \cdots & a_2 & a_1 & & a_{n+m-1} \\ a_{n+m-1} & a_{n+m-2} & a_{n+m-3} & & a_3 & a_2 & & a_1 \end{pmatrix}.$$

So the post process becomes,

$$n+m-1 \text{ bits}\left\{\begin{pmatrix} k_1 \\ \vdots \\ k_m \\ k_{m+1} \\ \vdots \\ k_{n+m-1} \end{pmatrix}\right. = k' = Ar'$$

$$= \begin{pmatrix} a_1 & a_{n+m-1} & a_{n+m-2} & & & a_4 & a_3 & a_2 \\ a_2 & a_1 & a_{n+m-1} & \cdots & & a_5 & a_4 & a_3 \\ a_3 & a_2 & a_1 & & & a_6 & a_5 & a_4 \\ & \vdots & & \ddots & & & \vdots & \\ a_{n+m-3} & a_{n+m-4} & a_{n+m-5} & & a_1 & a_{n+m-1} & & a_{n+m-2} \\ a_{n+m-2} & a_{n+m-3} & a_{n+m-4} & \cdots & a_2 & a_1 & & a_{n+m-1} \\ a_{n+m-1} & a_{n+m-2} & a_{n+m-3} & & a_3 & a_2 & & a_1 \end{pmatrix} \left.\begin{pmatrix} r_1 \\ r_2 \\ \vdots \\ r_n \\ 0 \\ 0 \\ \vdots \\ 0 \end{pmatrix}\right\} n+m-1 \text{ bits}.$$

We fill the $n$-bit-long vector $r'$ with $r = (r_1, r_2 \ldots r_n)$ and zeros behind it, while the final key sequence $k$ consists of the original $m$ bits of $k'$ whose exact location will be given later.

Although this $(n+m-1)\times(n+m-1)$ matrix is much larger and the efficiency seems to be lowed, we can avoid storage and computation of the direct matrix multiplication with the raw data vector: Noticing that for the $i$-th unit of $k$ we have $k_i = \sum_{j=1}^{n+m-1} a_{i-j+1(mod(n+m))} r_j$, which is exactly the model of discrete convolution. So it is reasonable to employ Fast Fourier Transformation (FFT) to accelerate the computational speed since the following definition and theorem guarantee the correctness of the acceleration.

**Definition 2:** The discrete convolution $z = x * y$ of two vector $x = (x_0, x_1 \dots x_{n-1})$ and $y = (y_0, y_1 \dots y_{n-1})$ is given as

$$z_i = \sum_{j=0}^{n-1} x_j y_{i-j mod n} \; 1 \le i \le n-1$$

**Theorem 1**: The discrete convolution of two vectors x and y has the following relation:

$$Fz = Fx \cdot Fy$$

Where $F$ means Fourier transformation which can be accelerated by FFT and $\cdot$ means dot product between two vectors[50].

By the theorem provided above, the matrix storage and direct multiplication are no longer necessary. We choose the random seed and the extended raw data provided above and use Fast Fourier Transformation to gain $Fa$ and $Fr'$ and the complexity of this step is $O(nlogn)$. In this step, we notice that there are only two vectors needed to store and compute. Then we dot product $Fa$ and $Fr'$ to get $Fk'$ and this step has complexity $O(n)$. Finally, we do inverse Fourier transformation so recovered $k'$ is on real field $\mathbb{R}$ and by round off and mod 2, the recovered $k'$ is on $\{0,1\}^{n+m-1}$.

Now, we need to get the same result as the Toeplitz multiplication. Since the required Toeplitz matrix is the submatrix of the last m rows and the first n column of the circulant matrix. So we fill $r_1$ to $r_n$ with raw numbers and $r_{n+1}$ to $r_{n+m-1}$ with 0, while preserving $k_n$ to $k_{n+m-1}$ as final key units. This result is equivalent to the original multiplication of Toeplitz matrix $T$ with raw data $r$. All of these calculations can be done in time complexity $O(n)$ so the overall complexity of the algorithm is determined by FFT and its inverse which are $O(nlogn)$.

In fact, the above analysis can be proved by the following theorem.

**Theorem 2**: A circulant matrix $A$ is unitarily similar to a diagonal matrix and the unitary matrix is $F = w^{ij}$, which is the representation of discrete Fourier transformation matrix. Furthermore, the elements of the diagonal matrix are equal to the elements of $Fa$. So in general, we have

$$FAF^{-1} = diag(Fa).$$

As a result,

$$k' = Ar' = F^{-1}diag(Fa)Fr' = F^{-1}(Fa \cdot Fr'),$$

where $F$ means Fourier transformation, and $F^{-1}$ means inverse Fourier transformation, and $\cdot$ means dot product[51].

### *C. Processing time comparison between Matrix Multiplication and FFT*

The time complexity of the algorithm is $O(n \log n)$[52], and when $n$ is large, post processing is significantly quicker compared with the original algorithm. We measured the average running time of these two algorithms under different raw data sequence lengths, $n$, shown in Fig. 4.

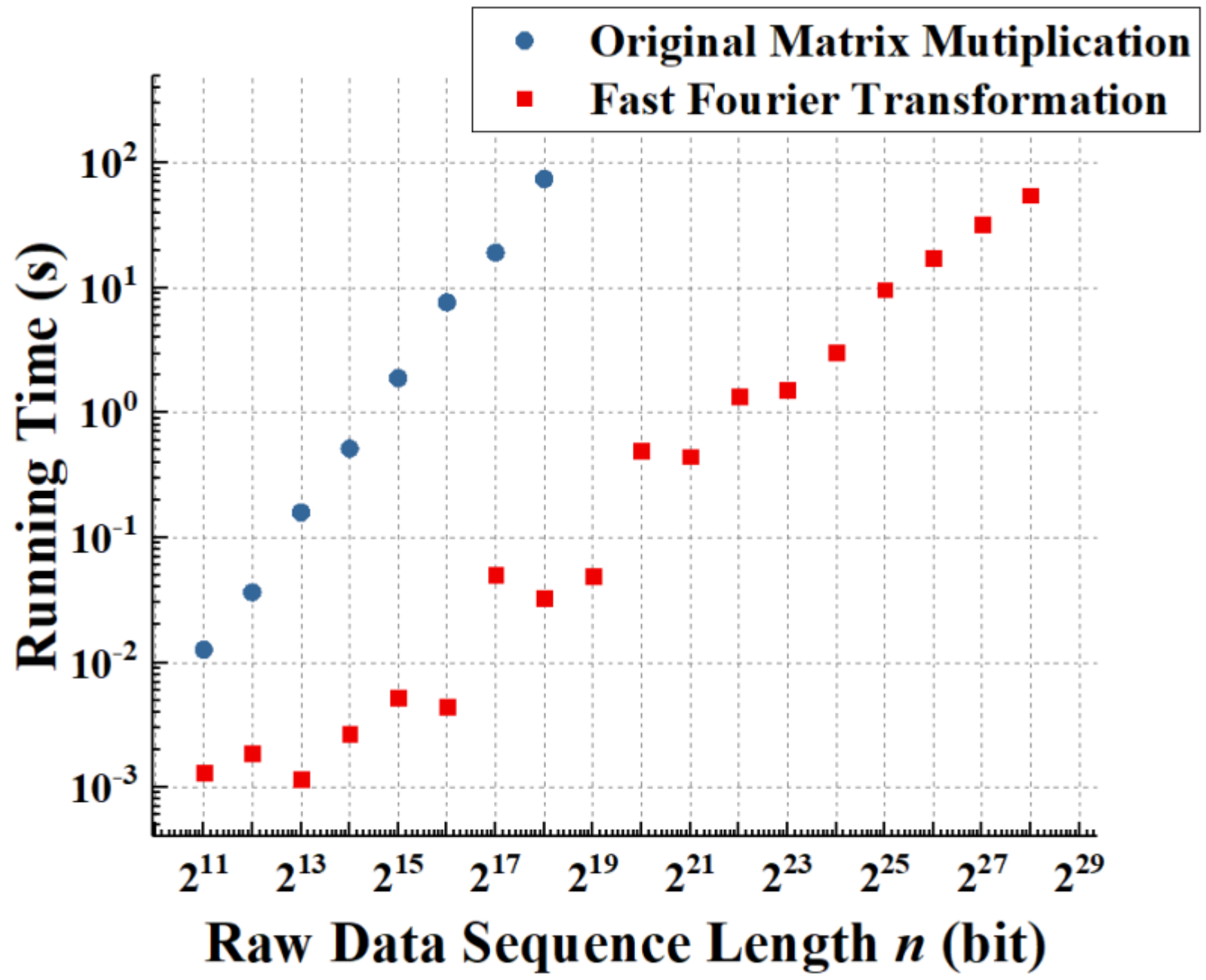

Fig. 4 Running time of two algorithms under different raw data sequence length $n$

From Table.1 and Fig. 4, we can see distinctly that FFT algorithm is much faster than original Toeplitz multiplication so the acceleration method is quite efficient. The running time of original algorithm rises continuously with $n$, but that of FFT algorithm fluctuates under some situations. We infer that this is related to specific calculation rules of FFT of the computers: When the size of input vector is $2^s$ for s is some integer, the calculation of FFT doesn't need divide and pad the vector to 2's power in process, therefore the practical calculation speed will be faster and its complexity will be much closer to the lower bound of $O(nlogn)$. As a result, it illustrates that we should find proper sequence length (or block length, see Sec. III. E) of processing every time when processing a very long raw data sequence, which will be discussed in the following section.

### *D. Processing with Modified Toeplitz Matrix*

The FFT algorithm has shown significant improvements on processing time. However, the seed length $n+m-1$ is still quite long and costs too much memory space and computational resource especially when the size of Toeplitz matrix is very large. Furthermore, since the random seed is composed of random numbers over $\{0,1\}$, a longer original random sequence is needed which demands a more powerful device. We resort to an improved way to accelerate the post-processing speed by reducing the seed length to almost a half, that is, we use a seed $\mathrm{a} = (a_1, a_2 \dots a_{n-1})$ as the seed the construct the modified Toeplitz matrix as follow[53]

$$A = \begin{pmatrix} a_{n-m} & a_{n-m-1} & a_{n-m-2} & & a_3 & a_2 & a_1 & 1 & 0 & 0 & & & & \\ a_{n-m+1} & a_{n-m} & a_{n-m-1} & \cdots & a_4 & a_3 & a_2 & 0 & 1 & 0 & \cdots & & 0 & \\ a_{n-m+2} & a_{n-m+1} & a_{n-m} & & a_5 & a_4 & a_3 & 0 & 0 & 1 & & & & \\ \vdots & & & \ddots & & \vdots & & & \vdots & & \ddots & & \vdots & \\ a_{n-3} & a_{n-4} & a_{n-5} & & a_m & a_{m-1} & a_{m-2} & & & & & 1 & 0 & 0 \\ a_{n-2} & a_{n-3} & a_{n-4} & \cdots & a_{m+1} & a_m & a_{m-1} & & 0 & & \cdots & 0 & 1 & 0 \\ a_{n-1} & a_{n-2} & a_{n-3} & & a_{m+2} & a_{m+1} & a_m & & & & & 0 & 0 & 1 \end{pmatrix}.$$

That is, we concatenate $A$ with an identical matrix $I_m$ to get $A_m = (A, I_m)$.

The modified Toeplitz matrix is still a family of universal hash functions which can also be used as randomness extractor to carry out post processing[53].

So we have $k = A_m r$:

$$m \text{ bits}\left\{\begin{pmatrix} k_1 \\ k_2 \\ \vdots \\ k_{m-1} \\ k_m \end{pmatrix}\right.$$

$$= \begin{pmatrix} a_{n-m} & a_{n-m-1} & a_{n-m-2} & & a_3 & a_2 & a_1 & 1 & 0 & 0 & & & & \\ a_{n-m+1} & a_{n-m} & a_{n-m-1} & \cdots & a_4 & a_3 & a_2 & 0 & 1 & 0 & \cdots & & 0 & \\ a_{n-m+2} & a_{n-m+1} & a_{n-m} & & a_5 & a_4 & a_3 & 0 & 0 & 1 & & & & \\ \vdots & & & \ddots & & \vdots & & & \vdots & & \ddots & & \vdots & \\ a_{n-3} & a_{n-4} & a_{n-5} & & a_m & a_{m-1} & a_{m-2} & & & & & 1 & 0 & 0 \\ a_{n-2} & a_{n-3} & a_{n-4} & \cdots & a_{m+1} & a_m & a_{m-1} & & 0 & & \cdots & 0 & 1 & 0 \\ a_{n-1} & a_{n-2} & a_{n-3} & & a_{m+2} & a_{m+1} & a_m & & & & & 0 & 0 & 1 \end{pmatrix} \left.\begin{pmatrix} r_1 \\ r_2 \\ \vdots \\ r_{n-1} \\ r_n \end{pmatrix}\right\} n \text{ bits},$$

As same as the previous process, a circulant matrix is created and the modification of the random seed and raw data are done. Here is the concrete way: Rather than transform $A_m$ into a circulant matrix, we only deal with the original matrix $A$ to make it circulant.

We construct a circulant matrix as follow:

$$A_c = \begin{pmatrix} a_1 & a_n & a_{n-1} & & a_4 & a_3 & a_2 \\ a_2 & a_1 & a_n & \cdots & a_5 & a_4 & a_3 \\ a_3 & a_2 & a_1 & & a_6 & a_5 & a_4 \\ & \vdots & & \ddots & & \vdots & \\ a_{n-2} & a_{n-3} & a_{n-4} & & a_1 & a_{n-1} & a_{n-2} \\ a_{n-1} & a_{n-2} & a_{n-3} & \cdots & a_2 & a_1 & a_{n-1} \\ a_n & a_{n-1} & a_{n-2} & & a_3 & a_2 & a_1 \end{pmatrix}$$

Here $a_n$ is a determined bit and we denote $a' = (a_1, a_2 \ldots a_{n-1}, a_n)$

We use the above circulant matrix to simplify the multiplication. In fact, since the last m rows and the first (n-m) columns of the $A_c$ are needed only, we should rule out the multiplication results of the added elements in circulant matrix. So we let the last m bit of raw data to be zero, that is $r' = (r_1, r_2 \ldots r_{n-m}, 0,0,0 \ldots 0)_n$, so it becomes

$$n \text{ bits} \left\{ k' = \begin{pmatrix} k'_1 \\ k'_2 \\ \vdots \\ k'_{n-1} \\ k'_n \end{pmatrix} \right.$$

$$= \begin{pmatrix} a_1 & a_n & a_{n-1} & & a_4 & a_3 & a_2 \\ a_2 & a_1 & a_n & \cdots & a_5 & a_4 & a_3 \\ a_3 & a_2 & a_1 & & a_6 & a_5 & a_4 \\ & \vdots & & \ddots & & \vdots & \\ a_{n-2} & a_{n-3} & a_{n-4} & & a_1 & a_{n-1} & a_{n-2} \\ a_{n-1} & a_{n-2} & a_{n-3} & \cdots & a_2 & a_1 & a_{n-1} \\ a_n & a_{n-1} & a_{n-2} & & a_3 & a_2 & a_1 \end{pmatrix} \left. \begin{pmatrix} r_1 \\ r_2 \\ \vdots \\ r_{n-m} \\ 0 \\ \vdots \\ 0 \end{pmatrix} \right\} n \text{ bits}$$

Then we use the FFT to accelerate the multiplication by calculating

$$k' = F^{-1}(Fa' \cdot Fr').$$

Finally, we modified the result of $k'$ to get $k$ by taking the multiplication of $I_m$ in to consideration. Since there is no information of the identical matrix $I_m$ in $A_c$, so $r_{i+1}$ is added as the result of the multiplication of $I_m$. We get the final k as

$$k_i = k'_i + r_{i+1} \ (n - m \leq i \leq n - 1)).$$

(We realize these post processing algorithms via Matlab coding, and all the codes is shown in Appendix.)

### *E. Proper Block Length in Processing a Fixed-length Long Raw Sequence*

When processing a raw random sequence, especially a long one, it is difficult for the computer to process it all at once, as its CPU can not read a vector with a huge length (such as 10Gbit), so we usually cut the sequence into several blocks and process them respectively (see Fig. 5). According to the results above, when FFT is used in post processing, processing time varies with the block size $n$. When processing a fixed-length sequence, there might be an optimized block length that ensures the shortest process time over the whole sequence.

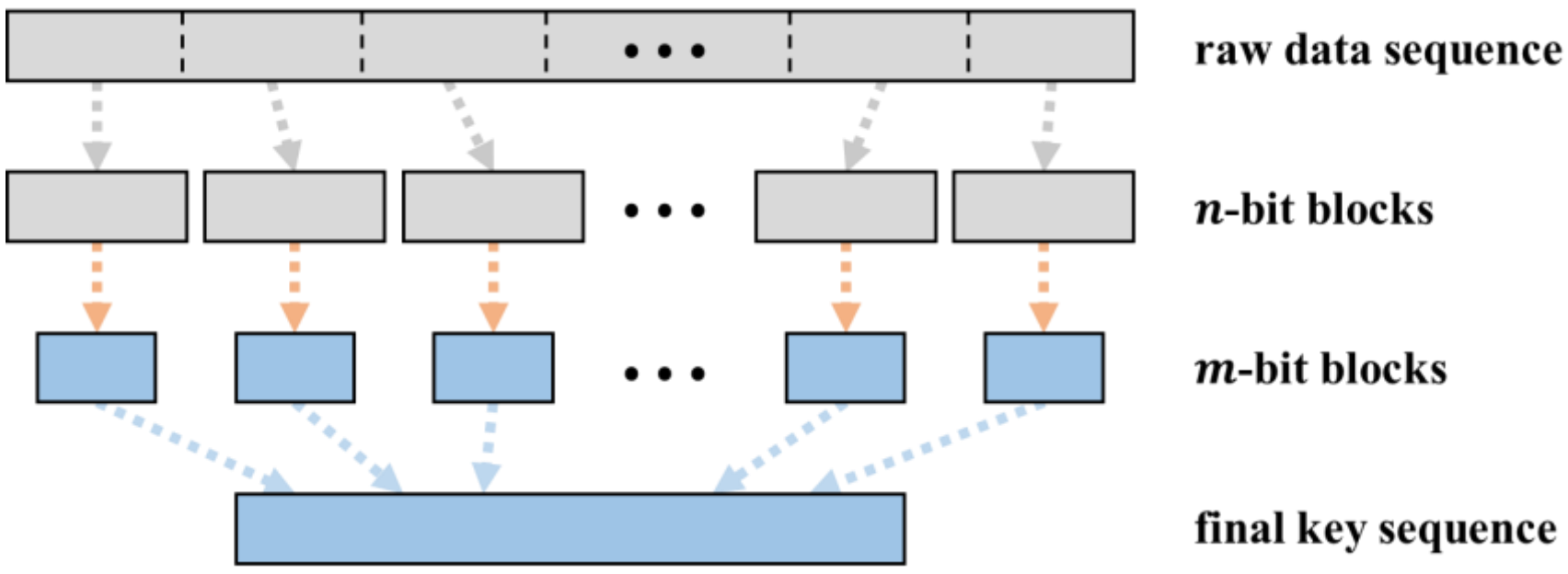

Fig. 5 Post processing on a long raw data sequence by split it into $n$-bit-long blocks and process every block respectively.

Final random numbers require not only to be process in short time. More importantly, final random numbers should be truly random. When the block is not long enough, the final random bits from every block maybe not random enough, because the distribution of a short sequence can be very different from uniform distribution random numbers. So after finding the optimized block length $n$, and processing raw data with $n$ to get the final random number sequence , it is necessary to test the sequence's randomness with standard test to guarantee that it is random and uniform distributed.

We choose a raw data of 10Gbit length, set different block length $n$ and measure the total process time $T$. Table 2 and Fig. 6 gives out the total processing time under different block lengths.

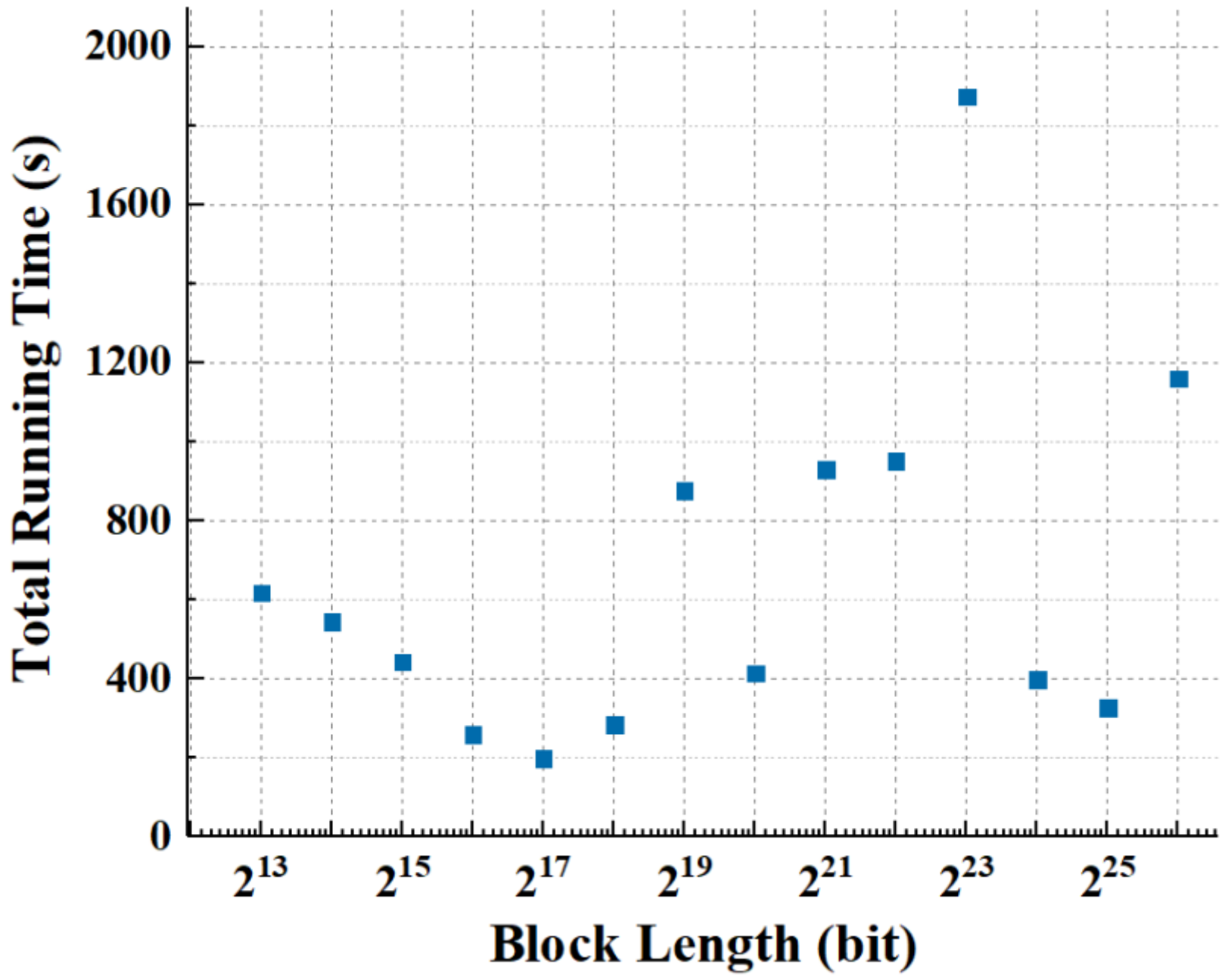

Fig. 6 Total processing time of 10G-bit-long raw sequence under different block lengths.

The result shows that there is a range for the shortest time $T$, appearing when $n = 2^{17}$, with the total time $T \approx 200\text{s}$.

Commonly, standard random test suites are adopted to test randomnesses, such as Diehard[54-55], TestU01[56] and NIST-STS suites[57]. We use NIST-STS test suite to test the randomness of our final sequence, and the final random sequence can pass all the tests verifying its randomness, shown in Fig. 7.

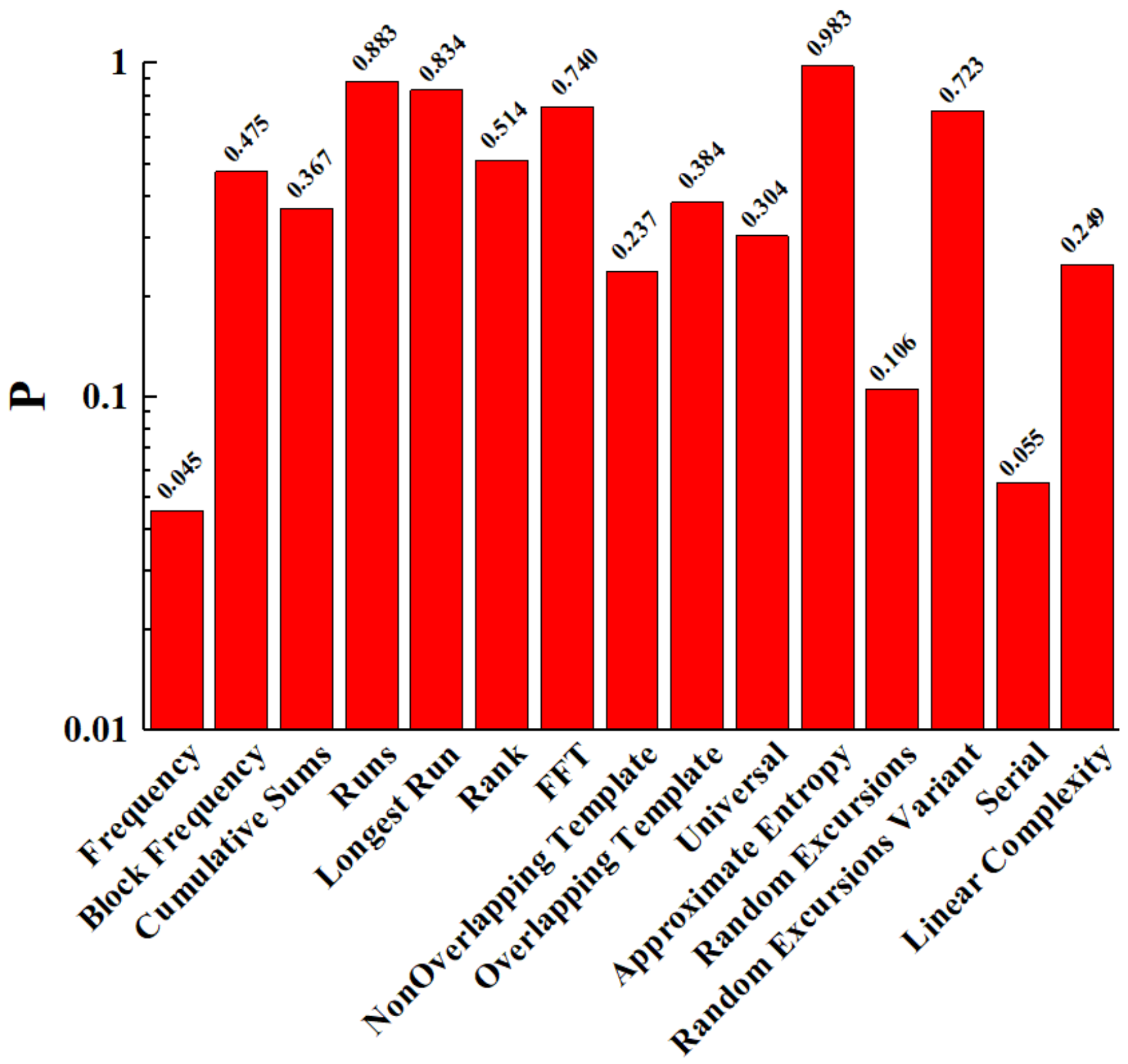

Fig. 7 Randomness test of the final random number sequence.

There are 15 independent tests in total in the NIST-STS test suites, and the result of every test is a confidence coefficient, or $P$ value. If the random sequence passes a test, the $P$ value of this test is between 0.01 and 0.99. Here in our experiment, the $P$ value of every test is between 0.01 and 0.99, indicating that the final random sequence can pass all the tests. Thus, the randomness of the whole 10Gbit-long random number sequence has been guaranteed, when we choose the block length $n = 2^{13}$.

## IV. Conclusion

In this paper, we discuss the evaluation of random sequences' min-entropy in practical QRNG when there is an eavesdropper, especially in measuring the laser's phase noise, and constructing an attack method for the eavesdropper to get information from the generator without being noticed. It can be achieved by changing the amplification coefficient of the photodiode detector and add noise in it. To account for this attack, which the same variance and mean value are constructed with a different class of distribution whose difference on statistical distance is not discernable, we propose to modify the calculation of min-entropy in practical QRNG, by finding an upper bound of mutual information between the generator and the eavesdropper, and minus this modification term $\Delta H_m$ from the original min-entropy to get the amount of secure raw data we can get from the generator.

This work proves there might be security loopholes in practical QRNG devices, and the eavesdropper can attack the generators by aiming at these loopholes. So the modification on min-entropy is essential for the generators' security, since it can prevent information leakage by extract less bits from raw data and suppress the mutual information $I(G{:}E)$, thus the eavesdropper will get no information about the final random numbers.

We also accelerate post processing procedure by using Fast Fourier transformation in the algorithm. It can improve the algorithm's time complexity from $O(mn)$ to $O(nlogn)$, and we achieve this algorithm in Matlab code. Also, Running time of post processing under different block lengths $n$ has been compared, whose result gives us a reference for choosing the proper block length in post processing. So, under the condition of guaranteed security, the randomness extractor has been realized more efficiently and passes the examination of NIST-STS suites.

In future work, we can construct more attack schemes, especially aiming at QRNG with different quantum randomness sources. This can indicate the insecurity in practical QRNG, urging us to find more solutions to close the security loopholes in actual devices. Meanwhile, a more precise way to evaluate the amount of randomness that we can extract from a quantum source is worth researching on, which can help modify or even replace min-entropy as an accurate solution. This may be reached with more useful tools in quantum information, such as Bell test[58] or device-independent QRNG[59-60].

As for the future work on post processing, we can also devise a theoretical upper bound of the leakage information in post processing. Moreover, a comparison of calculation speed between methods based on Toeplitz Matrix and Modified Toeplitz

Matrix (which are both accelerated by FFT) can be done to figure out how much time the modified Toeplitz matrix can save by reducing the seed length.

## Acknowledgement

We acknowledge Prof. Xiongfeng Ma and Mr. Hongyi Zhou for their great support on the research idea and experiment. We also acknowledge Mr. Daniel Comber Todd and Mr. Deion Hawkins for their kindly help on writing this article.

# Appendix

We realize post processing via Matlab coding, and the codes of different algorithms are as followed.

1. Original Toeplitz matrix:

```
n = 2^15;
m = gamma*n;
L = n+m-1;
RawData = Raw_Data_File(1:n);
MatrixSeed = Matrix_Seed_File(1:L);
MatrixSeed1 = [fliplr(MatrixSeed) fliplr(MatrixSeed)];
For i = 1:n
   Key1(i) = mod(dot(RawData,MatrixSeed1(n+m-i:2*n+m-i-1)),2);
end
```

2. Original Toeplitz matrix with FFT acceleration:

```
n = 2^15;
m = gamma*n;
L = n+m-1;
RawData = zeros(1,L);
RawData(1:n) = Raw_Data_File(1:n);
MatrixSeed2 = Matrix_Seed_File(1:L);
Key = mod(round(ifft((fft(MatrixSeed2).*fft(RawData)))),2);
Key2 = Key(1:m);
```

3. Simplified Toeplitz matrix with FFT acceleration:

```
n = 2^15;
m = gamma*n;
RawData1 = zeros(1,n);
RawData1(1:n-m) = Raw_Data_File(1:n-m);
```

```
RawData2 = Raw_Data_File(n-m+1:n);
MatrixSeed3 = Matrix_Seed_File(1:n);
Key = mod(round(ifft((fft(MatrixSeed3).*fft(RawData1)))),2);
Key3 = Key(1:m)+RawData2;
```